\documentclass[aps,pre,twocolumn]{revtex4}

\usepackage[version=3]{mhchem}
\usepackage{amsmath,amssymb}
\usepackage{graphicx}
\usepackage{color}
\usepackage{bm}

\begin{document}

\title{Convective Replica-Exchange in Ergodic Regimes}

\author{Giorgio F. Signorini$^\dag$}
\author{Edoardo Giovannelli$^\dag$}
\author{Yannick G. Spill$^\ddag$}
\author{Michael Nilges$^\ddag$}
\author{Riccardo Chelli$^\dag$}
\email{riccardo.chelli@unifi.it}
\affiliation{\ddag Institut Pasteur, Structural Bioinformatics and
Chemistry Department, 25 Rue du Docteur Roux, 75015 Paris, France}
\affiliation{\dag Dipartimento di Chimica, Universit\`a di Firenze,
  Via della Lastruccia 3, I-50019 Sesto Fiorentino, Italy \\ and European
  Laboratory for Non-linear Spectroscopy (LENS), Via Nello Carrara 1,
  I-50019 Sesto Fiorentino, Italy}

\date{\today}

\begin{abstract}
In a recent article (J. Comput. Chem. 2013, 34, 132-140), convective
Replica-Exchange (convective-RE) has been presented as an alternative
to the standard even-odd transition scheme. Computations on systems of
various complexity have shown that convective-RE may increase the
number of replica round-trips in temperature space with respect to the
standard exchange scheme, leading to a more effective sampling of
energy basins. Moreover, it has been shown that the method may prevent
the formation of bottlenecks in the diffusive walk of replicas through
the space of temperature states. By using an ideal temperature-RE
model and a classical harmonic-oscillator RE scheme, we study the
performances of convective-RE when ergodicity is not broken and
convergence of acceptance probabilities is attained. In this dynamic
regime, the round-trip ratio between convective and standard-RE is at
maximum $\sim 1.5$, a value much smaller than that observed in
non-ergodic simulations. For large acceptance probabilities, the
standard-RE outperforms convective-RE. Our observations suggest that
convective-RE can safely be used in either ergodic or non ergodic
regimes; however, convective-RE is advantageous only when bottlenecks
occur in the state-space diffusion of replicas, or when acceptance
probabilities are globally low. We also show that decoupling of the
state-space dynamics of the stick replica from the dynamics of the
remaining replicas improves the efficiency of convective-RE at low
acceptance probability regimes.
\end{abstract}

\maketitle


\section{Introduction}
\label{sec:introduction}

Convective-RE\cite{spill2013} is a simulation method designed to
enhance replica round-trip rate and to avoid the formation of replica
diffusion bottlenecks through the states of a generalized
ensemble\cite{okamoto2004, sugita1999, marinari1992,
  lyubartsev1992}. To achieve these benefits, convective-RE
artificially forces each replica to perform round-trips through the
states of the generalized ensemble. In doing so, global balance holds,
and a stationary distribution can be reached. For the sake of clarity
and to introduce a few key concepts, we will summarize the algorithm
underlying the convective-RE by using the words of Spill and
coworkers\cite{spill2013}.

{\em In the RE method, $N$ simulations of the same system are
  performed in parallel. Each simulation can be run in different
  ensembles} [also called states--Ed.]{\em, for example at different
  temperatures, ... which are all controlled via a set of control
  parameters. Each state will be given a unique label for quick
  reference. For simplicity, the states' labels are a monotonous
  function of the control parameters (in the canonical case, the
  lowest temperature is assigned label 1 and the highest is given
  label $N$). At a given exchange rate, adjacent states are allowed to
  exchange their conformations with a certain probability, given by
  the Metropolis criterion\cite{metropolis1953}. We call a given
  conformation whose simulation temperature is a function of time a
  replica. Let $S_i$ be the function that gives the state of replica
  $i$.}

{\em ... The convective algorithm is constructed as follows. Before
  trying the first exchange, a replica is chosen at random; we will
  refer to it as the stick replica; other replicas are passive. Let
  $i$ be the index of this stick replica, which is thus in state $S_i$
  at time 0. Upon the next exchange attempt, the transition matrix is
  chosen so as to allow the stick replica to exchange with its right
  neighbor state, $S_i + 1$. If the exchange fails, the simulation of
  each replica is continued. The next exchange attempt is however
  performed using the same transition matrix. The transition matrix is
  not changed until the exchange with the stick replica is accepted;
  the stick replica eventually is in state $S_i + 1$. The exchange
  matrix is then changed to allow for an exchange between $S_i + 1$
  and $S_i + 2$, and the same procedure is applied until the stick
  replica is in state $N$. At that point, the direction is reversed,
  and the transition matrices are chosen so that the stick replica can
  only go to lower state indices. Finally, when the stick replica
  reaches the lowest temperature state, the direction is reversed
  again and the same procedure is applied until the stick replica
  reaches its initial state $S_i$. The stick replica has then
  accomplished a round-trip in state space.  Then another replica,
  $j$, is chosen to be the stick replica and the same procedure is
  applied in turn until all replicas have been convective once, after
  which the stick replica is again replica $i$, $j$, and so forth.}

Convective-RE converges to the desired distribution. In the original
article\cite{spill2013} it is proven both analytically and numerically
that convective-RE satisfies global balance for the Boltzmann
distribution. Also, it is easily seen that in this method, like in any
replica-exchange method with $N$ states, there is a nonzero
probability of reaching any state from any other state after $N-1$
moves, so the sampling is regular. These two conditions
assure\cite{manousiouthakis1999} that there is a unique stationary
limit to the Markov chain, which is the Boltzmann distribution.

The ability of convective-RE to explore conformational space was
tested in systems of different degrees of complexity: alanine
dipeptide in implicit solvent, GB1 $\beta$-hairpin in explicit solvent
and the A$\beta_{25-35}$ homotrimer in a coarse grained
representation\cite{spill2013}. Comparison of convective-RE with the
standard method, namely the deterministic even-odd exchange
scheme\cite{lingenheil2009}, revealed that convective-RE significantly
enhances sampling of energy landscapes, and increases the number of
replica round-trips through temperature space. Sampling efficiency and
number of replica round-trips are known to be
correlated\cite{nadler2008, nadler2007, katzgraber2006, trebst2006} in
generalized ensemble\cite{park2007, chelli2010} approaches such as RE,
and increasing the number of round-trips is often a means of enhancing
exploration of conformational space. Given that convective-RE forces
(stick) replicas to perform round-trips through state space, it seems
surprising that replica round-trip rates increase in GB1
$\beta$-hairpin and A$\beta_{25-35}$ systems while they decrease in
the less complex alanine dipeptide system\cite{spill2013}. In fact,
the ratio between convective and standard-RE round-trip rates goes
from 48 and 8 for GB1 $\beta$-hairpin and A$\beta_{25-35}$ systems,
respectively, to 0.65 for the alanine dipeptide. In this case standard
method even outperforms convective-RE. Although these observations
point to some dependence of the performances of convective-RE on the
complexity of the system, a precise rationalization is difficult owing
to the non-ergodic regime to which GB1 $\beta$-hairpin and
A$\beta_{25-35}$ systems are subjected. In the latter systems,
ergodicity breaking is revealed by a couple of observations: (i) the
number of new structures detected during standard and convective-RE
simulations are still increasing at the end of the simulation (see
Fig. 3 of Ref. \cite{spill2013}) and (ii) in the convective-RE
simulation, an anomalous correlation between average potential
energies and populations of conformational basins has been observed
(see Table 1 and related discussion of Ref. \cite{spill2013}).

{\tiny
\begin{table*}
\begin{tabular}{l l l c}
\hline \hline
{\bf set} & $M =$ number of states (replicas) \\
{\bf set} &  $s =$ stick replica \\
{\bf loop} & for $t = 1, 2, ..., T$  & ($T =$ n. of tried exchanges needed \\
           &                           &  to obtain a round-trip of replica $s$) \\
           & {\bf set} ~ allpairs$(i) = 1$ ~ $\forall$ $i = 1, ..., M-1$ &  ($i \equiv$ pairs of states) \\
           & {\bf set} ~ xchpairs$(i) = 0$ ~ $\forall$ $i = 1, ..., M-1$ &  ($i \equiv$ pairs of states) \\
           & {\bf set} ~ xchpairs$(n) = 1$     &  ($n =$ pair of states involved in the \\
           &                                   &  exchange of the stick replica $s$) \\
           & {\bf set} ~ allpairs$(n) = 0$ \\
           & {\bf set} ~ allpairs$(n-1) = 0$   & (if $n-1 \in$ state-pair domain) \\
           & {\bf set} ~ allpairs$(n+1) = 0$   & (if $n+1 \in$ state-pair domain) \\
           & {\bf while} allpairs$()$ not null: \\
           & {\bf pick} at random a pair of states $p$ \\
           & {\bf if} ~ allpairs$(p) = 0$, go to previous step \\
           & {\bf set} ~ xchpairs$(p) = 1$ \\
           & {\bf set} ~ allpairs$(p) = 0$ \\
           & {\bf set} ~ allpairs$(p-1) = 0$ & (if $p-1 \in$ state-pair domain) \\
           & {\bf set} ~ allpairs$(p+1) = 0$ & (if $p+1 \in$ state-pair domain) \\
           & {\bf end while} \\
           & {\bf make} replica exchanges & (Use only replicas located in states \\
           &                       &  such that xchpairs$()=1$) \\
{\bf end} \\
{\bf select} & another stick replica $s$ \\
{\bf start} & loop again \\
\hline \hline
\end{tabular}
\caption{Pseudo-code for the random convective-RE scheme.}
\label{tab:1}
\end{table*}
}

We notice that for the above cases, ergodicity breaking is not
intrinsic to the systems or somehow generated by a sort of
unsuitability of the simulation algorithm. Rather, it is a consequence
of the small simulation time with respect to the times that would be
necessary to get a completely convergent sampling. Furthermore, a
common drawback of RE simulations of complex systems lies in the fact
that conformations are not decorrelated between successive exchanges
(which happened every 3 to 6 ps in Ref. \cite{spill2013}). This
correlation may dramatically increase the time needed to reach
ergodicity in these systems. We note that the broken ergodicity in GB1
$\beta$-hairpin and A$\beta_{25-35}$ simulations, prevented from
getting sound quantitative evaluations of the performances of
convective-RE.

In the present article, we tackle the above aspect of the problem by
exploiting two benchmark systems presenting ergodic behavior, for
which convergence of the acceptance probabilities of replica exchanges
and number of replica round-trips is almost achieved. We also consider
a third case where a local bottleneck is present in the diffusive walk
of replicas through state space. In particular, we will explore how
the acceptance probability of replica exchanges, modulated by the
number of replicas, affects the number of replica round-trips in
convective-RE compared to the standard method.

Finally, we will show that it is possible to improve the efficiency of
convective-RE, by supplying the convective scheme with an algorithm
aimed at decoupling the dynamics of stick and passive replicas through
state space.

\section{Setup of benchmark simulations}
\label{sec:methods}

The first benchmark case consists of a series of ideal RE simulations
in temperature space in which the potential energies of each replica
are sampled according to a Gaussian probability:
\begin{equation}
\rho_i(E) = \frac{1}{T_i \sqrt{2 \pi C}} \exp \left[ - \frac{(E - C
    T_i)^2}{2 C T_i^2} \right],
\label{eq:e_distribution}
\end{equation}
where $E$ is the potential energy of the replica lying in the state at
temperature $T_i$ and $C$ is the system heat capacity, which we assume
to be constant\cite{paschek2007}. Note that in
Eq. \ref{eq:e_distribution}, $C$ denotes the (extensive) heat capacity
in units of Boltzmann constant $k_B$ and refers to the potential
energy part of the total energy. In order to get equal average
acceptance probabilities for the exchanges between neighboring
replicas, the spacing law of temperatures is\cite{okamoto1991}
\begin{equation}
T_i = T_{\rm min} \left( \frac{T_{\rm max}}{T_{\rm min}}
\right)^{\frac{i-1}{N-1}},
\label{eq:spacing_law}
\end{equation}
where $[T_{\rm min}, T_{\rm max}]$ is the temperature range covered by
the $N$ states/replicas of the generalized ensemble, with temperature
index $i=1, \dots, N$. For both convective and standard-RE algorithms
we have performed several RE simulations with a number of replicas $N$
ranging from 8 to 100, but with a fixed temperature range $T_{\rm min}
= T_1 = 300$ K to $T_{\rm max} = T_N = 1500$ K and with temperature
spacing according to Eq. \ref{eq:spacing_law}. In these simulations,
replica potential energies $E$ are drawn at each RE step from the
distributions given by Eq. \ref{eq:e_distribution}, with $C = 500$ for
the heat capacity. The Metropolis criterion\cite{metropolis1953} is
applied to evaluate the outcome of the exchange attempts. In order to
get convergent estimates of both acceptance probabilities and number
of replica round-trips, long simulations lasting 10$^7$ steps have
been carried out. Replica exchanges are attempted at every step.

\begin{figure}
\begin{center}
\includegraphics[width=1\columnwidth,keepaspectratio=true]{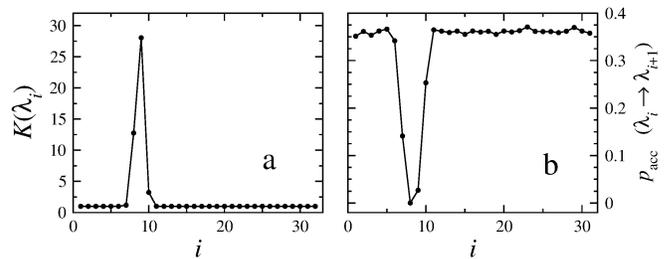}
\end{center}
\caption{Panel a: sequence of the $K(\lambda_i)$ parameter as a
  function of the index $i$ employed in a B-simulation with 32
  replicas. Panel b: acceptance probability $p_{\rm acc}$ of replica
  transitions $\lambda_i \rightarrow \lambda_{i+1}$ as a function of
  the index $i$ computed for a standard-RE B-simulation with 32
  replicas.}
\label{fig:acceptance_probability}
\end{figure}

The other benchmark case consists of RE simulations in which the
states are described by one-dimensional harmonic oscillators, whose
Hamiltonians depend parametrically by a factor $\lambda_i$ entering
both in the equilibrium position and in the force constant of the
oscillators: $H_i(x) = \frac{1}{2} K(\lambda_i) ( x - \lambda_i )^2$.
At each step, the replica coordinates, $x$, are picked according to a
Boltzmann distribution with $\beta^{-1} = k_BT = 1$:
\begin{equation}
\rho_i(x) = Z^{-1} \exp(-H_i(x)),
\label{eq:x_distribution}
\end{equation}
with $Z$ being the partition function of the system. In all
simulations, the parameter $\lambda_i$ ranges from $\lambda_1 = 0$ to
$\lambda_N = 40$ in equally spaced steps, {\em i.e.}, $\Delta \lambda
= \lambda_{i+1} - \lambda_i = 40/(N-1)$. Two series of simulations
have been performed with different definition of $K(\lambda_i)$:
A-simulations, in which $K(\lambda_i) = 1$ for all states $i = 1, 2,
\dots, N$; and B-simulations, in which
\begin{equation}
K(\lambda_i) = 1 + 30 ~ {\rm e}^{-( \lambda_i - 10 )^2}.
\label{eq:k_distribution}
\end{equation}
This choice for $K(\lambda_i)$ allows us to introduce a bottleneck in
correspondence of few states associated with $\lambda_i$ around the
value of 10. For example, in a B-simulation with 32 replicas,
Eq. \ref{eq:k_distribution} gives rise to the $K(\lambda_i)$ sequence
shown in Figure \ref{fig:acceptance_probability}a. In a standard
B-simulation, such a sequence leads to an acceptance probability
bottleneck mainly localized at the transitions $\lambda_8
\Leftrightarrow \lambda_9$ and $\lambda_9 \Leftrightarrow
\lambda_{10}$ (see Figure \ref{fig:acceptance_probability}b). For
these two transitions, the average acceptance probabilities are in
fact $2.4 \times 10^{-4}$ and $2.7 \times 10^{-2}$, respectively,
versus values of about 0.36 for replica transitions far from the
bottleneck.

Like in temperature-RE simulations, a series of A-simulations have
been performed with $N$ ranging from 8 to 100, each lasting $10^7$
steps. At every step, the replica coordinates $x$ are drawn according
to Eq. \ref{eq:x_distribution} and simultaneously a replica exchange
is attempted. B-simulations are instead $3.4 \times 10^5$ steps long,
with $N$ ranging from 32 to 96. The setup of B-simulations has been
devised to mimic a system where all parameters, {\em i.e.} $\Delta
\lambda$, $K(\lambda_i)$ and $N$, yield acceptance probabilities
comparable to those of typical atomistic simulations in the
generalized ensemble\cite{marsili2010, chelli2012, chelli2012b} (see
Figure \ref{fig:acceptance_probability}b). Also the number of steps
has been chosen to get simulation times consistent with
state-of-the-art atomistic simulations. For example, with a number of
attempted replica-exchanges of $3.4 \times 10^5$ and considering that
replica exchanges are typically performed every few ps in atomistic
simulations, our B-simulations would cover times of the order of
$10^2$-$10^3$ ns. In the specific case that replica exchanges are
attempted every 6 ps, as in A$\beta_{25-35}$ simulations of
Ref. \cite{spill2013}, our simulation time would correspond to
slightly more than 2 $\mu$s (versus 0.5 $\mu$s of A$\beta_{25-35}$
simulations). Since all above quantities take realistic values, we
expect that the number of round-trips is comparable to that of
atomistic simulations. On the other side, we notice that, owing to the
relatively small number of steps in B-simulations, the average
round-trip numbers may not be large enough to allow precise
evaluations. This aspect can be particularly critical in those
B-simulations that are affected by low acceptance probabilities around
the bottleneck. Therefore, in order to confirm our outcomes, we also
performed B-simulations with standard and convective-RE schemes
lasting $10^7$ steps, in which the number of round-trips is expected
to increase by a factor of about 30 with respect to B-simulations $3.4
\times 10^5$ steps long.

\section{Results and discussion}
\label{sec:results}

\begin{figure}
\begin{center}
\includegraphics[width=1\columnwidth,keepaspectratio=true]{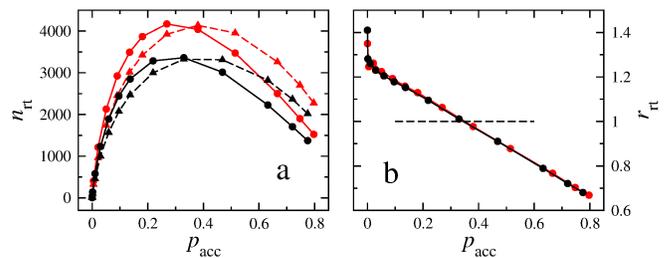}
\end{center}
\caption{Panel a: Number of round-trips per replica, $n_{\rm rt}$, as
  a function of the average acceptance probability, $p_{\rm acc}$,
  calculated from temperature-RE simulations and A-simulations (red
  and black colors, respectively) with various numbers of replicas
  (circles and triangles from left to right represent simulations
  performed with $N = 8, 10, 12, 14, 16, 18, 20, 24, 30, 40, 60, 80,
  100$). Simulations using the standard method are reported with
  triangles, while simulations using convective-RE are reported with
  circles. Panel b: Ratio between the numbers of replica round-trips
  estimated with convective-RE and standard simulations, $r_{\rm rt} =
  [n_{\rm rt}]_{\rm conv.} / [n_{\rm rt}]_{\rm stand.}$, in
  temperature-RE simulations and A-simulations (red and black colors,
  respectively). The dashed line indicates the crossover regime.}
\label{fig:set-a}
\end{figure}

In Figure \ref{fig:set-a}a we report the number of round-trips per
replica, $n_{\rm rt}$, as a function of the average acceptance
probability, $p_{\rm acc}$, obtained from standard and convective-RE
simulations performed in $\lambda$ space (A-simulations) and in
temperature space (temperature-RE simulations). The acceptance
probability has been modulated by varying the number $N$ of replicas
in the simulations. The trends of the curves obtained from temperature
and $\lambda$-space simulations are similar because the sampling of
potential energy and replica coordinates, respectively, occurs
according to Normal distributions (Eqs. \ref{eq:e_distribution} and
\ref{eq:x_distribution}). In convective-RE, the maximum performance is
reached at $p_{\rm acc} \simeq 0.28$, versus $p_{\rm acc} \simeq 0.40$
of standard-RE\cite{lingenheil2009}. It is worth noting that, at small
$p_{\rm acc}$ values, convective-RE outperforms standard-RE, whereas
the opposite occurs at large $p_{\rm acc}$. The crossover regime falls
at $p_{\rm acc} \simeq 0.35$ for both types of simulations.

The performances of convective and standard-RE can be compared by
plotting the ratio of the number of replica round-trips estimated by
the two methods, {\em i.e.}, $r_{\rm rt} = [n_{\rm rt}]_{\rm conv.} /
[n_{\rm rt}]_{\rm stand.}$, as a function of $p_{\rm acc}$. The plots
are shown in Figure \ref{fig:set-a}b. We note that, for temperature-RE
simulations in the regime of very small $p_{\rm acc}$ values ($\sim
3.2 \times 10^{-4}$), $n_{\rm rt}$ for convective-RE is 1.35 times
greater than $n_{\rm rt}$ for standard-RE. Similar outcomes are
obtained in $\lambda$-space simulations ($r_{\rm rt} \simeq 1.4$ at
$p_{\rm acc} \simeq 5.4 \times 10^{-5}$). In both cases, the ratio
$r_{\rm rt}$, though significantly greater than 1, is about one order
of magnitude smaller than that gained from non-ergodic simulations of
atomistic systems\cite{spill2013}. Interestingly, the standard method
appears to be more efficient than convective-RE when $p_{\rm acc}$ is
large, and reaches $r_{\rm rt} \simeq 0.67$ for $p_{\rm acc} \simeq
0.78$. These observations are roughly consistent with the outcomes of
Ref. \cite{spill2013}. In fact, under the well-grounded assumption
that crossover is not attained in the simulations of GB1
$\beta$-hairpin and A$\beta_{25-35}$\cite{note1}, we may argue that
$p_{\rm acc}$ is in average smaller for GB1 $\beta$-hairpin than for
A$\beta_{25-35}$ from the fact that the number of round-trips per
replica, {\em i.e.} $n_{\rm rt} / {\rm Simul.Time}$, is smaller in its
standard-RE simulation ($n_{\rm rt} / {\rm Simul.Time} = 3.1 \times
10^{-4}$ found in standard-RE simulation of GB1 $\beta$-hairpin,
versus $9.2 \times 10^{-3}$ found in standard-RE simulation of
A$\beta_{25-35}$). Thus, on the basis of our results, a greater
$r_{\rm rt}$ is expected for the GB1 $\beta$-hairpin
simulation. Indeed, this is in agreement with the outcomes of
Ref. \cite{spill2013} ($r_{\rm rt} = 48$ for GB1 $\beta$-hairpin
versus $r_{\rm rt} = 8$ for A$\beta_{25-35}$). Moreover, the much
larger number of round-trips per replica achieved in the standard-RE
simulation of alanine dipeptide points to a $p_{\rm acc}$ even greater
than that of the A$\beta_{25-35}$ simulation. Considering that $r_{\rm
  rt} = 0.65$, we may argue that, in the case of the alanine
dipeptide, crossover regime has been largely surpassed.

In spite of this qualitative agreement between the present study and
the results of Ref. \cite{spill2013}, we notice the very large
quantitative difference in the $r_{\rm rt}$ values. To explain such
differences one could suppose that the ratio $r_{\rm rt}$ goes
asymptotically to infinity as $p_{\rm acc}$ goes to zero, but this is
not strongly supported by the quite large $n_{\rm rt}$, and hence the
non-negligible values of $p_{\rm acc}$, observed in the simulations of
A$\beta_{25-35}$. On the other hand, such an asymptotic regime would
not be of great relevance in practice, because it would be reached
when the number of round-trips is too small to make the RE simulation
really effective. In fact, in the limit of zero $p_{\rm acc}$, the
number of round-trips per replica in convective-RE becomes negligible,
so that all benefits of the method would be lost. For example, the
number of round-trips per replica in convective-RE A-simulations
obtained by using 8 replicas is 6.9, while $r_{\rm rt}$ is only $\sim
1.4$. Probably, to obtain $r_{\rm rt}$ values of the order of 10,
simulations with 1 or less round-trips per replica should be
performed.

In light of these observations, and also considering that
convective-RE preserves ergodicity\cite{spill2013}, the outcomes of
Ref. \cite{spill2013} could be interpreted as due to the lack of
convergence in the computed $n_{\rm rt}$. On the other side, the
relatively high number of round-trips observed especially in the
A$\beta_{25-35}$ simulation leaves open the question about the
achievement of convergence. Some answer to this problem could be
obtained by increasing enormously the simulation time and decreasing
the rate of replica-transition attempts as well, which is however out
of the reach of current computational resources.

\begin{figure}
\begin{center}
\includegraphics[width=1\columnwidth,keepaspectratio=true]{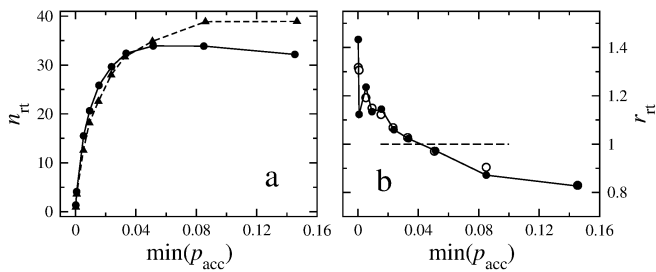}
\end{center}
\caption{Panel a: Number of round-trips per replica, $n_{\rm rt}$, as
  a function of the minimum acceptance probability, $\min(p_{\rm
    acc})$, calculated from B-simulations with various numbers of
  replicas (from left to right, data of simulations with $N = 32, 38,
  48, 52, 56, 60, 64, 70, 80, 96$ are reported). B-simulations using
  the standard method are reported with triangles, while B-simulations
  using convective-RE are reported with circles. Panel b: Ratio
  between the numbers of replica round-trips estimated with
  convective-RE and standard B-simulations, $r_{\rm rt} = [n_{\rm
      rt}]_{\rm conv.} / [n_{\rm rt}]_{\rm stand.}$, as a function of
  $\min(p_{\rm acc})$ (filled circles). The $r_{\rm rt}$ quantity
  calculated from B-simulations of $10^7$ steps is reported for
  comparison (open circles). The dashed line indicates the crossover
  regime.}
\label{fig:bottleneck}
\end{figure}

The relative performance of standard and convective-RE in the presence
of a bottleneck in replica-transitions can be appreciated in Figure
\ref{fig:bottleneck}a, where we report $n_{\rm rt}$ as a function of
the minimum acceptance probability among all replica transitions,
$\min(p_{\rm acc})$, computed for B-simulations. We point out that, in
the presence of a bottleneck in the replica transitions, the overall
diffusion of the replicas through state space is regulated by the
lowest acceptance probability occurring at bottleneck transition
itself. In this situation, $\min(p_{\rm acc})$ rather than the $p_{\rm
  acc}$ (which is averaged over all replica transitions) becomes more
appropriate to monitor the $n_{\rm rt}$ trend. The behavior of the
curves of Figure \ref{fig:bottleneck}a looks like that of Figure
\ref{fig:set-a}a, though the crossover regime is moved down to
$\min(p_{\rm acc}) \simeq 3.3 \times 10^{-2}$ (see
Ref. \cite{note2}). Similarly, the performances of the convective-RE
relative to the standard method can be better appreciated from the
$r_{\rm rt}$ vs. $\min(p_{\rm acc})$ plot reported in Figure
\ref{fig:bottleneck}b. At small values of $\min(p_{\rm acc})$, the
curve shows a quite noisy behavior due to poor convergence arising
from the small number of simulation steps. As a matter of fact,
increasing the number of steps from $3.4 \times 10^5$ to $10^7$, a
more regular, but substantially identical, trend is observed (open
circles in Figure \ref{fig:bottleneck}b). The value of $r_{\rm rt}$
ranges from $\sim 1.4$ at $\min(p_{\rm acc}) \simeq 2.2 \times
10^{-4}$ ($\sim 1.3$, in better convergence conditions) to $\sim 0.83$
at $\min(p_{\rm acc}) \simeq 0.15$ (practically unchanged, in better
convergence conditions). With respect to the uniform distribution of
acceptance probabilities enforced in A-simulations, the presence of a
bottleneck does not change significantly the performances of
convective-RE relative to the standard method. Convective-RE is better
than standard-RE when low acceptance probabilities occur, due to a
nonuniform $p_{\rm acc}$ distribution featured by the presence of a
bottleneck\cite{spill2013} (see Figure
\ref{fig:acceptance_probability}b).

However, small acceptance probabilities imply low round-trip rates
also in convective-RE, which may prevent the RE scheme from being
effective. For example, in B-simulations, the best performances of
convective-RE with respect to standard-RE ($r_{\rm rt} \simeq 1.4$) is
obtained with 32 replicas, yielding $\min(p_{\rm acc})=2.2 \times
10^{-4}$. Unfortunately, this acceptance probability leads to a small
number of round-trips ($n_{\rm rt} \simeq 1.3$), which can typically
be increased by changing the number of replicas. In our case, to
obtain a workable $n_{\rm rt}$ value, {\em e.g.} $n_{\rm rt} \simeq
15$, we would have needed to increase the number of replicas to
48. With this number of replicas, the ratio $r_{\rm rt}$ lowers to
$\sim 1.2$. A further increase of $N$, {\em e.g.} $N = 64$, yields a
satisfactory number of round-trips ($n_{\rm rt} \simeq 32$), but the
advantages of convective-RE almost disappear ($r_{\rm rt} \simeq
1$). In summary, by doubling the number of replicas from 32 to 64, the
round-trip rate increases by more than 20 with a practically complete
loss of the advantages of convective-RE with respect to standard RE.

\begin{figure}
\begin{center}
\includegraphics[width=0.8\columnwidth,keepaspectratio=true]{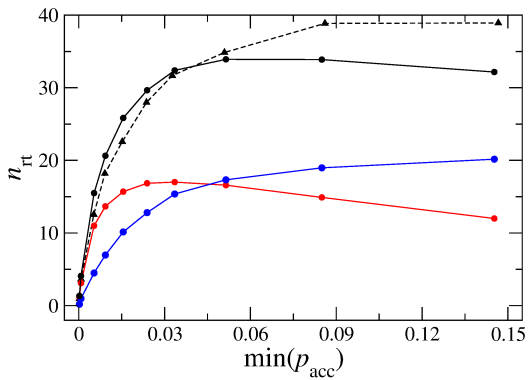}
\end{center}
\caption{Number of round-trips per replica, $n_{\rm rt}$, as a
  function of the minimum acceptance probability, $\min(p_{\rm acc})$,
  calculated from B-simulations ($3.4 \times 10^5$ steps) with various
  numbers of replicas (from left to right data of simulations with $N
  = 32, 38, 48, 52, 56, 60, 64, 70, 80, 96$ are reported). Simulations
  using the standard method are reported with triangles. For
  convective-RE simulations, the total $n_{\rm rt}$, and the
  contributions from stick and passive replicas are shown (black, red
  and blue circles, respectively). The data obtained with
  B-simulations of $10^7$ steps differ from those reported here by a
  factor of $\sim 30$, but the overall behavior remains practically
  unchanged (data available upon request).}
\label{fig:stick-passive}
\end{figure}

Another interesting aspect of convective-RE observed in
A$\beta_{25-35}$ simulation\cite{spill2013} was the almost unexpected
distribution of the number of round-trips between stick and passive
replicas. Among the 706 round-trips globally observed in the
A$\beta_{25-35}$ simulation, 435 were realized by stick replicas,
while the remaining 271 round-trips were accomplished by passive
replicas. Interestingly, this number is still three times greater than
in the standard simulation, during which only 88 round-trips were
counted. This fact was explained by observing that replica exchanges
are correlated in convective-RE: ``{\em When the stick replica crosses
  the bottleneck, a passive replica crosses it as well, but in the
  other direction.}''. This behavior is not confirmed by the current
B-simulations, as shown in Figure \ref{fig:stick-passive}, where we
report $n_{\rm rt}$ as a function of $\min(p_{\rm acc})$ for the
standard and convective-RE, detailing the contributions to the total
number of round-trips from stick and passive replicas. We note that,
at variance with the A$\beta_{25-35}$ simulation data, at higher
values of $\min(p_{\rm acc})$ the contribution of passive replicas to
$n_{\rm rt}$ is greater than that of the stick replica; also, that
each contribution is smaller than the number of round-trips in the
standard-RE.

\section{Combining convective-RE with a random pair-replica selection scheme}
\label{sec:random-convective-RE}

The efficiency of convective-RE with respect to the standard even-odd
scheme stems from generating replica exchanges aimed at moving a
single replica, the so-called stick replica, along one of the two
possible directions in state space. When the upper (or lower) end
state is reached, exchanges are attempted to guide the stick replica
in the opposite direction, {\em i.e.}, towards the other end
state. This process is repeated until the stick replica completes a
round-trip. During this forced walk, the other replicas, called
passive replicas, are ``constrained'' to move according to the stick
replica, on the basis of an even-odd scheme. When the acceptance
probabilities are globally large, the constrained motion of passive
replicas prevents their free diffusion through the states, eventually
leading to a significant reduction of the number of round-trips
(Figures \ref{fig:set-a} and \ref{fig:bottleneck}). This constrained
motion of passive replicas has instead no dramatic effects as
bottlenecks occur, because, in such a case, the dynamics of replicas
is dominated by the convective component of the motion. However, in
the presence of bottlenecks, we expect that constraining passive
replicas to the stick one, somehow slows down the walk diffusion of
the former. In fact, passive replicas will continue to swap between
two states, without a resultant net diffusion, until the stick replica
will overcome the bottleneck.

Based on the above observations, we envisage a possibility of
improving the diffusion of passive replicas through the space of
states by decoupling their dynamics from that of the stick
replica. This can be realized by supplying the convective (even-odd)
scheme with a stochastic criterion to choose the replica pairs which
must undergo attempted exchanges. This strategy basically allows
decoupling between passive and stick replicas, so that the exchanges
involving the passive replicas are independent of each other and,
especially, independent of the stick replica. A pseudo-code related to
this random-convective exchange scheme is reported in Table
\ref{tab:1}.

\begin{figure}
\begin{center}
\includegraphics[width=1.0\columnwidth,keepaspectratio=true]{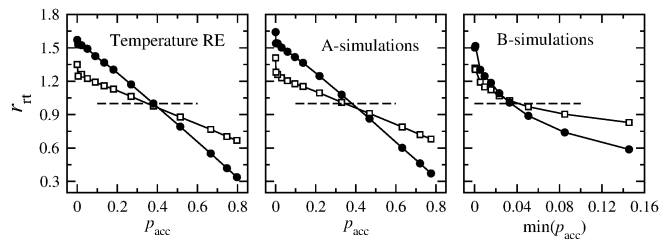}
\end{center}
\caption{Ratio between the numbers of replica round-trips estimated
  with convective and standard RE simulations, $r_{\rm rt} = [n_{\rm
      rt}]_{\rm conv.} / [n_{\rm rt}]_{\rm stand.}$, as a function of
  $p_{\rm acc}$ (temperature-RE and A-type simulations) and
  $\min(p_{\rm acc})$ (B-simulations). In each panel, the even-odd and
  random convective schemes are compared (squares and circles,
  respectively). The dashed line indicates the crossover regime. For
  B-simulations, results obtained with $10^7$ steps are reported.}
\label{fig:random-convective-RE}
\end{figure}

The round-trip efficiency ratios, $r_{\rm rt}$ (see above for
definition), obtained from temperature-RE simulations, A-simulations
and B-simulations using the random-convective scheme are compared in
Figure \ref{fig:random-convective-RE} with the convective even-odd
approach\cite{spill2013}. Overall, we note that, when convective and
random-pair selection schemes are combined, $r_{\rm rt}$ increases,
but only in the regime of low acceptance probability. In this regime,
the increase of efficiency of the random-convective scheme with
respect to the convective even-odd method, is relevant in percentage,
even if it appears quite modest in absolute terms. The opposite trend,
that is a decrease of efficiency of the random convective-RE, is
observed for large acceptance probabilities.

In light of the previous discussion, the former behavior is not
surprising. On the contrary, the loss of performance of the random
convective-RE at large acceptance probabilities has a more subtle
origin, which can be understood by imagining the performances of the
three algorithms, standard even-odd, convective even-odd and
random-convective, in the limit of $p_{\rm acc} \sim 1$. Under this
assumption, using the standard even-odd as well as the convective
even-odd schemes leads to a very fast round-trip rate. In such cases,
in fact, a round-trip takes exactly $2 N$ attempted replica exchanges,
where $N$ is the number of states in the simulation. Note that this is
the maximum efficiency one can get from RE simulations in terms of
round-trip rate. These performances cannot clearly be obtained by
using the random-convective scheme, because the random choice of pairs
of replicas to be exchanged can easily invert their walk in the space
of states. Consistently with the present results, it is worth noting
that loss of efficiency was also observed in the so-called stochastic
even-odd scheme\cite{lingenheil2009}, which has strong analogies with
the random-convective sequence of Table \ref{tab:1}.

\section{Concluding remarks}
\label{sec:conclusions}

In this article, we have investigated the performances of the
convective transition scheme\cite{spill2013} for RE simulations
relative to the standard even-odd deterministic approach. Calculations
have been carried out in ideal fully ergodic conditions by means of
toy-model simulations. Overall, our calculations show that, when
ergodicity and convergence are both in place, the performances of
convective-RE, in terms of replica round-trip rate, do not exceed 1.5
times the ones obtained with the standard replica-transition
scheme. Above some crossover value of the acceptance probability,
$p_{\rm acc}$, which occurs below the optimal $p_{\rm acc}$
value\cite{note3} in simulations with uniform $p_{\rm acc}$
distribution, standard even-odd algorithm starts to be competitive
with convective-RE, outperforming the latter as large $p_{\rm acc}$s
are attained. Similar results are obtained if a bottleneck in replica
transitions is present. These observations are consistent with the
results reported in Ref. \cite{spill2013}, though the performances of
convective-RE seem to be much less striking than those observed in
more complex systems. Although the true reasons of these discrepancies
are not completely understood, we believe that uncertainties in
round-trip rates due to the loss of ergodicity arising from the
shortness of the simulations (with respect to the dynamics needed to
get effective sampling), may play some role. Therefore, it would be
interesting to evaluate the performances of convective-RE in complex
systems when round-trip rate reaches satisfactory convergence, which
may occur only by performing microsecond scale simulations.

Furthermore, an attempt at improving the performances of convective-RE
has been done by devising a stochastic scheme to select the passive
replica pairs undergoing exchanges. The exchange mechanism, based on
decoupling the diffusion motions of stick and passive replicas through
the state space, proved to be effective with respect to the even-odd
convective scheme as bottlenecks are present or the acceptance
probabilities are globally low. Conversely, the stochastic selection
criterion makes the efficiency significantly worse at high acceptance
probability regimes.

In summary, we have shown that, when bottlenecks occur in RE
simulations due to low acceptance probabilities arising, {\em e.g.},
from a not optimized spacing between the states, or when the
acceptance probabilities are globally low due to small numbers of
states/replicas, the use of even-odd convective-RE or random
convective-RE schemes leads to improvements in the round-trip
rate. The benefits of convective schemes are lost if the acceptance
probabilities are globally large.

\begin{acknowledgments}
We thank Piero Procacci for stimulating discussions and Pierluigi
Cresci for technical support. This work was supported by European
Union Contract RII3-CT-2003-506350. Y. G. Spill and M. Nilges would
like to acknowledge the ERC grant FP7-IDEAS-ERC 294809.
\end{acknowledgments}



\begin{thebibliography}{22}
\expandafter\ifx\csname natexlab\endcsname\relax\def\natexlab#1{#1}\fi
\expandafter\ifx\csname bibnamefont\endcsname\relax
  \def\bibnamefont#1{#1}\fi
\expandafter\ifx\csname bibfnamefont\endcsname\relax
  \def\bibfnamefont#1{#1}\fi
\expandafter\ifx\csname citenamefont\endcsname\relax
  \def\citenamefont#1{#1}\fi
\expandafter\ifx\csname url\endcsname\relax
  \def\url#1{\texttt{#1}}\fi
\expandafter\ifx\csname urlprefix\endcsname\relax\def\urlprefix{URL }\fi
\providecommand{\bibinfo}[2]{#2}
\providecommand{\eprint}[2][]{\url{#2}}

\bibitem[{\citenamefont{Spill et~al.}(2013)\citenamefont{Spill, Bouvier, and
  Nilges}}]{spill2013}
\bibinfo{author}{\bibfnamefont{Y.~G.} \bibnamefont{Spill}},
  \bibinfo{author}{\bibfnamefont{G.}~\bibnamefont{Bouvier}}, \bibnamefont{and}
  \bibinfo{author}{\bibfnamefont{M.}~\bibnamefont{Nilges}},
  \bibinfo{journal}{J. Comput. Chem.} \textbf{\bibinfo{volume}{34}},
  \bibinfo{pages}{132} (\bibinfo{year}{2013}).

\bibitem[{\citenamefont{Okamoto}(2004)}]{okamoto2004}
\bibinfo{author}{\bibfnamefont{Y.}~\bibnamefont{Okamoto}}, \bibinfo{journal}{J.
  Mol. Graphics Modell.} \textbf{\bibinfo{volume}{22}}, \bibinfo{pages}{425}
  (\bibinfo{year}{2004}).

\bibitem[{\citenamefont{Sugita and Okamoto}(1999)}]{sugita1999}
\bibinfo{author}{\bibfnamefont{Y.}~\bibnamefont{Sugita}} \bibnamefont{and}
  \bibinfo{author}{\bibfnamefont{Y.}~\bibnamefont{Okamoto}},
  \bibinfo{journal}{Chem. Phys. Lett.} \textbf{\bibinfo{volume}{314}},
  \bibinfo{pages}{141} (\bibinfo{year}{1999}).

\bibitem[{\citenamefont{Marinari and Parisi}(1992)}]{marinari1992}
\bibinfo{author}{\bibfnamefont{E.}~\bibnamefont{Marinari}} \bibnamefont{and}
  \bibinfo{author}{\bibfnamefont{G.}~\bibnamefont{Parisi}},
  \bibinfo{journal}{Europhys. Lett.} \textbf{\bibinfo{volume}{19}},
  \bibinfo{pages}{451} (\bibinfo{year}{1992}).

\bibitem[{\citenamefont{Lyubartsev et~al.}(1992)\citenamefont{Lyubartsev,
  Martsinovski, Shevkunov, and Vorontsov-Velyaminov}}]{lyubartsev1992}
\bibinfo{author}{\bibfnamefont{A.~P.} \bibnamefont{Lyubartsev}},
  \bibinfo{author}{\bibfnamefont{A.~A.} \bibnamefont{Martsinovski}},
  \bibinfo{author}{\bibfnamefont{S.~V.} \bibnamefont{Shevkunov}},
  \bibnamefont{and} \bibinfo{author}{\bibfnamefont{P.~N.}
  \bibnamefont{Vorontsov-Velyaminov}}, \bibinfo{journal}{J. Chem. Phys.}
  \textbf{\bibinfo{volume}{96}}, \bibinfo{pages}{1776} (\bibinfo{year}{1992}).

\bibitem[{\citenamefont{Metropolis et~al.}(1953)\citenamefont{Metropolis,
  Rosenbluth, Rosenbluth, Teller, and Teller}}]{metropolis1953}
\bibinfo{author}{\bibfnamefont{N.}~\bibnamefont{Metropolis}},
  \bibinfo{author}{\bibfnamefont{A.~W.} \bibnamefont{Rosenbluth}},
  \bibinfo{author}{\bibfnamefont{M.~N.} \bibnamefont{Rosenbluth}},
  \bibinfo{author}{\bibfnamefont{A.~H.} \bibnamefont{Teller}},
  \bibnamefont{and} \bibinfo{author}{\bibfnamefont{E.}~\bibnamefont{Teller}},
  \bibinfo{journal}{J. Chem. Phys.} \textbf{\bibinfo{volume}{21}},
  \bibinfo{pages}{1087} (\bibinfo{year}{1953}).

\bibitem[{\citenamefont{Manousiouthakis and Deem}(1999)}]{manousiouthakis1999}
\bibinfo{author}{\bibfnamefont{V.~I.} \bibnamefont{Manousiouthakis}}
  \bibnamefont{and} \bibinfo{author}{\bibfnamefont{M.~W.} \bibnamefont{Deem}},
  \bibinfo{journal}{J. Chem. Phys.} \textbf{\bibinfo{volume}{110}},
  \bibinfo{pages}{2753} (\bibinfo{year}{1999}).

\bibitem[{\citenamefont{Lingenheil et~al.}(2009)\citenamefont{Lingenheil,
  Denschlag, Mathias, and Tavan}}]{lingenheil2009}
\bibinfo{author}{\bibfnamefont{M.}~\bibnamefont{Lingenheil}},
  \bibinfo{author}{\bibfnamefont{R.}~\bibnamefont{Denschlag}},
  \bibinfo{author}{\bibfnamefont{G.}~\bibnamefont{Mathias}}, \bibnamefont{and}
  \bibinfo{author}{\bibfnamefont{P.}~\bibnamefont{Tavan}},
  \bibinfo{journal}{Chem. Phys. Lett.} \textbf{\bibinfo{volume}{478}},
  \bibinfo{pages}{80} (\bibinfo{year}{2009}).

\bibitem[{\citenamefont{Nadler et~al.}(2008)\citenamefont{Nadler, Meinke, and
  Hansmann}}]{nadler2008}
\bibinfo{author}{\bibfnamefont{W.}~\bibnamefont{Nadler}},
  \bibinfo{author}{\bibfnamefont{J.~H.} \bibnamefont{Meinke}},
  \bibnamefont{and} \bibinfo{author}{\bibfnamefont{U.~H.~E.}
  \bibnamefont{Hansmann}}, \bibinfo{journal}{Phys. Rev. E}
  \textbf{\bibinfo{volume}{78}}, \bibinfo{pages}{061905}
  (\bibinfo{year}{2008}).

\bibitem[{\citenamefont{Nadler and Hansmann}(2007)}]{nadler2007}
\bibinfo{author}{\bibfnamefont{W.}~\bibnamefont{Nadler}} \bibnamefont{and}
  \bibinfo{author}{\bibfnamefont{U.~H.~E.} \bibnamefont{Hansmann}},
  \bibinfo{journal}{Phys. Rev. E} \textbf{\bibinfo{volume}{75}},
  \bibinfo{pages}{026109} (\bibinfo{year}{2007}).

\bibitem[{\citenamefont{Katzgraber et~al.}(2006)\citenamefont{Katzgraber,
  Trebst, Huse, and Troyer}}]{katzgraber2006}
\bibinfo{author}{\bibfnamefont{H.~G.} \bibnamefont{Katzgraber}},
  \bibinfo{author}{\bibfnamefont{S.}~\bibnamefont{Trebst}},
  \bibinfo{author}{\bibfnamefont{D.~A.} \bibnamefont{Huse}}, \bibnamefont{and}
  \bibinfo{author}{\bibfnamefont{M.}~\bibnamefont{Troyer}},
  \bibinfo{journal}{J. Stat. Mech.} \textbf{\bibinfo{volume}{3}},
  \bibinfo{pages}{03018} (\bibinfo{year}{2006}).

\bibitem[{\citenamefont{Trebst et~al.}(2006)\citenamefont{Trebst, Troyer, and
  Hansmann}}]{trebst2006}
\bibinfo{author}{\bibfnamefont{S.}~\bibnamefont{Trebst}},
  \bibinfo{author}{\bibfnamefont{M.}~\bibnamefont{Troyer}}, \bibnamefont{and}
  \bibinfo{author}{\bibfnamefont{U.~H.~E.} \bibnamefont{Hansmann}},
  \bibinfo{journal}{J. Chem. Phys.} \textbf{\bibinfo{volume}{124}},
  \bibinfo{pages}{174903} (\bibinfo{year}{2006}).

\bibitem[{\citenamefont{Park and Pande}(2007)}]{park2007}
\bibinfo{author}{\bibfnamefont{S.}~\bibnamefont{Park}} \bibnamefont{and}
  \bibinfo{author}{\bibfnamefont{V.~S.} \bibnamefont{Pande}},
  \bibinfo{journal}{Phys. Rev. E} \textbf{\bibinfo{volume}{76}},
  \bibinfo{pages}{016703} (\bibinfo{year}{2007}).

\bibitem[{\citenamefont{Chelli}(2010)}]{chelli2010}
\bibinfo{author}{\bibfnamefont{R.}~\bibnamefont{Chelli}}, \bibinfo{journal}{J.
  Chem. Theory Comput.} \textbf{\bibinfo{volume}{6}}, \bibinfo{pages}{1935}
  (\bibinfo{year}{2010}).

\bibitem[{\citenamefont{Paschek et~al.}(2007)\citenamefont{Paschek, Nymeyer,
  and Garcia}}]{paschek2007}
\bibinfo{author}{\bibfnamefont{B.}~\bibnamefont{Paschek}},
  \bibinfo{author}{\bibfnamefont{H.}~\bibnamefont{Nymeyer}}, \bibnamefont{and}
  \bibinfo{author}{\bibfnamefont{A.~E.} \bibnamefont{Garcia}},
  \bibinfo{journal}{J. Struct. Biol.} \textbf{\bibinfo{volume}{157}},
  \bibinfo{pages}{524} (\bibinfo{year}{2007}).

\bibitem[{\citenamefont{Okamoto et~al.}(1991)\citenamefont{Okamoto, Fukugita,
  Nakazawa, and Kawai}}]{okamoto1991}
\bibinfo{author}{\bibfnamefont{Y.}~\bibnamefont{Okamoto}},
  \bibinfo{author}{\bibfnamefont{M.}~\bibnamefont{Fukugita}},
  \bibinfo{author}{\bibfnamefont{T.}~\bibnamefont{Nakazawa}}, \bibnamefont{and}
  \bibinfo{author}{\bibfnamefont{H.}~\bibnamefont{Kawai}},
  \bibinfo{journal}{Protein Eng.} \textbf{\bibinfo{volume}{4}},
  \bibinfo{pages}{639} (\bibinfo{year}{1991}).

\bibitem[{\citenamefont{Marsili et~al.}(2010)\citenamefont{Marsili, Signorini,
  Chelli, Marchi, and Procacci}}]{marsili2010}
\bibinfo{author}{\bibfnamefont{S.}~\bibnamefont{Marsili}},
  \bibinfo{author}{\bibfnamefont{G.~F.} \bibnamefont{Signorini}},
  \bibinfo{author}{\bibfnamefont{R.}~\bibnamefont{Chelli}},
  \bibinfo{author}{\bibfnamefont{M.}~\bibnamefont{Marchi}}, \bibnamefont{and}
  \bibinfo{author}{\bibfnamefont{P.}~\bibnamefont{Procacci}},
  \bibinfo{journal}{J. Comput. Chem.} \textbf{\bibinfo{volume}{31}},
  \bibinfo{pages}{1106} (\bibinfo{year}{2010}).

\bibitem[{\citenamefont{Chelli and Signorini}(2012{\natexlab{a}})}]{chelli2012}
\bibinfo{author}{\bibfnamefont{R.}~\bibnamefont{Chelli}} \bibnamefont{and}
  \bibinfo{author}{\bibfnamefont{G.~F.} \bibnamefont{Signorini}},
  \bibinfo{journal}{J. Chem. Theory Comput.} \textbf{\bibinfo{volume}{8}},
  \bibinfo{pages}{830} (\bibinfo{year}{2012}{\natexlab{a}}).

\bibitem[{\citenamefont{Chelli and
  Signorini}(2012{\natexlab{b}})}]{chelli2012b}
\bibinfo{author}{\bibfnamefont{R.}~\bibnamefont{Chelli}} \bibnamefont{and}
  \bibinfo{author}{\bibfnamefont{G.~F.} \bibnamefont{Signorini}},
  \bibinfo{journal}{J. Chem. Theory Comput.} \textbf{\bibinfo{volume}{8}},
  \bibinfo{pages}{2552} (\bibinfo{year}{2012}{\natexlab{b}}).

\bibitem[{not({\natexlab{a}})}]{note1}
\bibinfo{note}{In this case the occurrence of a crossover should be evaluated
  considering the plot of $n_{\rm rt}$ {\em versus} the smallest $p_{\rm acc}$,
  because $n_{\rm rt}$ would depend basically on the replica transition
  $\lambda_i \Leftrightarrow \lambda_{i+i}$ with lowest probability.}

\bibitem[{not({\natexlab{b}})}]{note2}
\bibinfo{note}{A comparison of B-simulations with A-simulations is however
  improper because in the former case acceptance probabilities are not
  distributed uniformly.}

\bibitem[{not({\natexlab{c}})}]{note3}
\bibinfo{note}{In the present context, the optimal $p_{\rm acc}$ is the one for
  which the number of round-trips is maximum}.

\end{thebibliography}
\end{document}